


\font\twelvecaps = cmcsc10 scaled\magstep1
\font\sixrm = cmr6
\font\sixi  = cmmi6
\def\oneonehalfspace {%
   \lineskip        .20ex
   \baselineskip    4.0ex
   \lineskiplimit     0ex
   \parskip          .80ex plus .04ex minus .20ex
}%
\def\dline{\noalign{\hrule height1pt\vskip 2pt\hrule height 1pt\vskip1pt}}
\def\cline{\noalign{\vskip 3pt\hrule height 1pt\vbox to 5pt{}}}
\def\bline{\noalign{\vskip 5pt\hrule height1pt}}
\def\yyskip{\penalty-100\vskip6pt plus6pt minus4pt}
\def\yskip{\penalty-50\vskip3pt plus3pt minus2pt}
\def\sr#1{_{\hbox{\sixrm {#1}}}}
\def\si#1{_{\hbox{\sixi {#1}}}}
\def\eb#1{\hbox {\bf {#1}}}
\def\er#1{\hbox {\rm {#1}}}
\def\pp{\par\noindent\hangindent 0.4in \hangafter 1}
\def\lapprox{
     \mathrel{
           \vcenter{
                    \offinterlineskip \hbox {$<$}
                    \kern 0.3ex \hbox{$\sim$}
}}}
\def\gapprox{
     \mathrel{
           \vcenter{
                    \offinterlineskip \hbox {$>$}
                    \kern 0.3ex \hbox{$\sim$}
}}}
\def\ric{R\si{IC}}
\def\tc{T\si{IC}}
\def\tic8{T\si{IC8}}
\oneonehalfspace
\rm

\centerline {\bf Accretion Disk Coronae in High Luminosity Systems}
\vskip .3truein
\centerline {\bf Stephen D. Murray{$^{1,2}$}, John I. Castor{$^{2}$},}
\centerline {\bf Richard I. Klein{$^{1,2}$}, \& Christopher F. McKee{$^{1,3}$}}
\vskip .3truein

\noindent{$^1$}Dept. of Astronomy, 601 Campbell Hall, Univ. of California,
Berkeley, CA  94720

\noindent{$^2$}Lawrence Livermore Nat'l Lab., L-58, P.O. Box 808, Livermore,
CA, 94550

\noindent{$^3$}Dept. of Physics, Univ. of California, Berkeley, CA  94720

\vskip 0.3truein

\centerline{\bf ABSTRACT}

We present the results of self-consistent models of Compton-heated accretion
disk coronae.  The models are calculated using a new method for computing
monochromatic radiative transfer in two-dimensions.  The method splits the
radiation into direct and scattered components.  The direct radiation is
computed by calculating the optical depth along rays, while transfer of the
scattered radiation is approximated by flux-limited diffusion.  The resulting
code agrees with more accurate treatments to within 50\%, and
is highly efficient, making it practical for use in
large hydrodynamic simulations.  The coronal models are used to confirm the
results of earlier work, and to extend it to higher luminosities.  In contrast
to earlier work, which found the outer disks to be shadowed by the inner
corona at high luminosities, we find our results to form an almost
continuous extension of the models at lower luminosities.  This is due to
the presence of multiply-scattered radiation, which acts to partially offset
the loss of direct radiation from the central source.  Although the analytic
methods derived at lower luminosities cannot be used to derive the coronal
structure for $L/L\sr{Edd}\gapprox0.1$, the results of the models are
amenable to semiempirical fits.  We also discuss possible observational
consequences of the results for coronal veiling and line fluorescence from the
disk.

\vskip 0.15in
\noindent To appear in {\it The Astrophysical Journal}

\vfill\eject

\centerline{\bf 1. INTRODUCTION}
\yyskip

Systems such as X-ray binaries and active galactic nuclei are generally
observed to be X-ray sources.  If the accretion disks in these systems are
exposed to the X-rays from either the central source or inner accretion disk,
then a Compton-heated corona is expected to naturally result (Shakura \&
Sunyaev 1973).  Scattering
of X-rays from the central source by the corona may then increase the
irradiation of the outer regions of the accretion disk, with several
important consequences.  Heating by the X-rays may lead to winds, driving
substantial mass loss (Begelman, McKee, \& Shields 1983; Woods et al. 1994).
The irradiation may also exceed the viscous energy output of the outer
disk.  In this case, the internal disk structure may be significantly
affected, in such a way as to alter or even remove the thermal instability
responsible for driving the outbursts of cataclysmic variables, some X-ray
binaries, and possibly active galactic nuclei (Tuchman, Mineshige, \& Wheeler
1990).

The coronal structure is also crucial in determining the shape of the X-ray
emission lines used to infer the temperature and velocities in the accretion
disks.  Using simple assumptions about the structure of accretion disk coronae,
Kallman (1990), and Ko \& Kallman (1993), find that the lack of emission
near 6.4~keV in the iron K line seen in X-ray binaries implies the absence
of irradiation of the outer accretion disk.  This may imply that the outer
disk is shadowed by the inner coronal gas, a conclusion supported by the
models of London (1985), though these had convergence problems.  The
models also made several simplifying
assumptions which may have affected the results.  Most importantly, the
transfer of scattered radiation was assumed to occur purely vertically
throughout the corona, and radiation from the accretion disk itself was
ignored.

Ostriker, McKee, \& Klein (1991; hereafter OMK) examined systems in which the
scattered radiation was assumed to be dominated by singly-scattered photons.
While this limited their models to systems with luminosities
$L\lapprox0.1L\sr{Edd}$, they were able to
treat the radiative transfer accurately in two-dimensions.  They also included
the effects of radiation from the accretion disk.  Because this radiation
has a lower characteristic temperature than that from the central source, its
effect is to cool the corona, reducing the scale height and optical
depths.  They found that the outer disk was strongly irradiated, with possibly
significant observable effects.

In this contribution, we perform self-consistent models of Compton-heated
accretion disk coronae.  The assumptions made are similar to those of OMK.
An important difference is in the treatment of the radiation field.  While
our method of radiative transfer is more approximate, it includes
multiply-scattered radiation.  We are therefore able to extend the models
of OMK to much higher luminosities, and examine the effects of shadowing by
the optically-thick inner corona upon the outer regions.  As in the previous
work, the assumption of hydrostatic equilibrium limits the models to the
inner part of the disk.  For example, for a disk in a binary X-ray source,
our analysis is limited to radii $\lapprox4\times10^9$~cm.
To simulate models at larger radii, where winds lead to
substantial mass loss, requires that the radiative transfer be coupled
to hydrodynamics.  The great computational efficiency of the radiative
transfer method presented here makes this possible, and will allow us to
explore the many dynamic effects of accretion in systems with luminosities
$L\approx L\sr{Edd}$ in future work.

We proceed in \S~2 by describing the numerical method used in our radiative
transfer calculations, and comparing it with the results of more detailed
transfer calculations.  In \S~3, we use the method to construct self-consistent
models of accretion disk coronae.  Where appropriate, these are compared with
the results of earlier work, while we also extend the range of parameters
for which models are computed.  In \S~4, we discuss possible observational
consequences of the coronal models.  Finally, in \S~5 we summarize our
conclusions.

\yyskip
\centerline{\bf 2. NUMERICAL METHOD}
\yyskip

\centerline{\rm 2.1. Considerations}
\yyskip

We need to calculate radiative transfer in two-dimensions in an
axisymmetric geometry.  For photoionized gas at coronal temperatures, the
pure absorption opaticy is dominated by bound-free transitions of iron for
which the mean opacity is typically an order of magnitude smaller than the
scattering opacity.  We can therefore approximate the opacity as being due
only to pure electron scattering.  The optical depths are sufficiently low
that the Thomson limit applies, which allows us to consider
monochromatic transfer.  The primary difficulty in the calculations is in the
optical depths,
which vary greatly and are strongly anisotropic, with the optical depth
measured parallel to the accretion disk greatly exceeding that measured
in the perpendicular direction.  The method to be used must also be highly
efficient, due to the long computation times already required by the
two-dimensional, time-dependent hydrodynamic code with which it will be joined
(Woods et al. 1994).

We use the geometrically-corrected dimensionless mean intensity
$$f_i\equiv{{4\pi J_i}\over{L/4\pi R^2}},\eqno{(2.1)}$$
where $J_i$ is the mean intensity of radiation source $i$, and $R$ is the
distance from the central source, which has luminosity
$L$.  Our method divides the radiation into direct and scattered
components,
denoted $f\sr{dir}$ and $f\sr{sc}$ respectively.  Splitting the radiation
field in this way has several advantages.  Most importantly, it improves
the accuracy with which the optically thick and thin limits are treated.
In the former, $f\sr{sc}\gg f\sr{dir}$, and the flux-limited
diffusion used to compute $f\sr{sc}$ forms an excellent approximation.
In the latter, $f\sr{sc}\ll f\sr{dir}\equiv e^{-\tau}$, where the
optical depth $\tau$ is calculated
very accurately along rays, as described below.  In addition,
the direct radiation can be weighted as a function of ray angle,
allowing the use of analytic results for the central regions, such as
those of OMK to increase accuracy in simulations where the central
region may not be entirely resolved.

\yyskip
\centerline{\rm 2.2. Direct Radiation}
\yyskip

The direct radiation is computed by calculating $\tau$ along rays from
the central source to each cell in the computational grid.  An extremely
rapid method is to store the identification
and path lengths of the cells through which each ray passes in a
three-dimensional array.  This information needs to be calculated only
once for a fixed grid, and can then be accessed to calculate $\tau$ using
updated values
of the opacities.  The resultant algorithm vectorizes quite easily.
Unfortunately, while it is extremely quick, this approach requires a very large
amount of memory, making it
impractical for grids containing more than about 10$^4$ cells.

For larger grids, an algorithm is needed which calculates the ray information
each time, in a way that vectorizes and so is as economical as possible.
Let the indices of $R$ and $Z$ on the grid be $j$ and $k$, respectively.
We begin by finding the value of $Z$ corresponding to each gridline
$R=R(j)$
crossed by the ray traveling from the central source to the cell of
interest.  In the case of a uniform grid, the corresponding value of $k$ of
the cell into which the ray is passing is then given by the integer operation
$$k = {\rm int}(Z/\Delta Z)+1,\eqno{(2.2)},$$
where $\Delta Z$ is the cell size in the axial direction.  A uniform grid is
not required, and $\Delta Z$ can be a function of $Z$ so
long as the above operation can be written
in an analytic form.  Thus, for example, $\Delta Z$ may increase
geometrically, as it does for the results presented in \S~3.  The same
operation is then performed to compute
the values of $R$ and $j$ for each gridline $Z=Z(k)$ crossed by the ray.  The
resulting
values of $R$, $Z$, $j$, and $k$ are then stored in one-dimensional arrays,
indexed by $i=j+k-1$, such that they are automatically ordered.
The optical depths are then calculated for each ray as
$$\tau=\sum\limits_{i=1}^{i_{max}}\chi_i\Delta r_i,\eqno{(2.3)}$$
where $\Delta r_i=\left(\delta R_i^2+\delta Z_i^2\right)^{1/2}$, $\delta R_i$
and $\delta Z_i$ are the radial and axial distances, respectively, traversed by
the ray through each cell, and $\chi$ is the opacity.

In cases where the lower boundary of the grid corresponds to the surface
of a flared disk (see below, and Woods et al. 1994), then the rectangular
computational grid corresponds, in real space, to a curved grid, in which
$Z(k)$ is a function of $R$, ie.
$$Z\sr{true}(k)=Z\sr{comp}(k)+Z\sr{off}(R),\eqno{(2.4)}$$
where $Z\sr{off}(R)$ is the $Z$ offset of the lower boundary.  In this
case, the rays are curved on the computational grid, and the values of $R$
are double-valued for each $Z(k)$, although the higher value may
be off the grid.  To find $R$ we must now solve the transcendental equation
$$Z(k)=mR-Z\sr{off}(R),\eqno{(2.5)}$$
where $m$ is the slope of the ray in real space.
In order to maintain vectorization,
equation~(2.5) is solved by a Newton-Raphson method with a fixed number of
iterations.  If the higher root corresponds to a value of $R$ less than that
of the zone of interest, then
both roots are solved for, and stored.  Otherwise, only the lower root is of
interest.

\yyskip
\centerline{\rm 2.3. Scattered Radiation}
\yyskip

In the absence of time-dependence, and velocity-dependent effects,
the flux of scattered radiation is given by the first moment equation
$$\nabla\cdot{\eb F}\sr{sc}=c\chi E\sr{dir},\eqno{(2.6)}$$
where $E\sr{dir}$ is the energy density of the direct radiation.  To compute
{\bf F}$\sr{sc}$, we use the Flux-Limited Diffusion (FLD) approximation
(Levermore \& Pomraning 1981)
$${\eb F\sr{sc}}=-{c\over\chi}D\si{F}\nabla E\sr{sc},\eqno{(2.7)}$$
where $E\sr{sc}$ is the energy density of scattered radiation, and $D\si{F}$
is given by the rational approximation
$$D\si{F}={{\lambda({\rm X})}\over\omega},\eqno{(2.8)}$$
$$\lambda({\rm X})={{2+{\rm X}}\over{6+3{\rm X}+{\rm X}^2}},\eqno{(2.9)}$$
with
$${\rm X}=-{\vert{\nabla E\sr{sc}\vert}\over{\chi\omega E\sr{sc}}}.
\eqno{(2.10)}$$
For pure Thomson scattering the effective albedo, $\omega=1$.  Equation~(2.7)
is nonlinear in the sense that $D_F$ is a function of $E\sr{sc}$.  This is
dealt with by using the previous estimate of $E\sr{sc}$ in computing
$D_F$, and iterating.  Only two or three iterations are found to be necessary
for the resulting energy densities to converge to within a percent.

The above equations are solved iteratively using the Incomplete
Cholesky-Conjugate Gradient (ICCG) Method (Kershaw 1978).  This method is
robust and vectorizes well.  While the method solves the full two-dimensional
system implicitly, and so does not require operator-splitting, the fluxes
at each cell face are computed using the corresponding derivatives in
equations~(2.7) and (2.10), as would be done in an operator-split method
(Bowers \& Wilson 1991).  For example, in computing the flux in the $R$
direction, $\nabla E\si{sc}$ is replaced by ${{\partial E\si{sc}}
\over{\partial R}}$.  This is
done so as to provide the symmetric matrix required by the ICCG method.

\yyskip
\centerline{\rm 2.4. Tests}
\yyskip

As a test of the above method, we have compared its output with that of
the two-dimensional version of the radiative code ALTAIR, which solves the
full transport equations for line transfer using a tensor moment approach,
taking into account the full angular variation of the radiation field
(Klein et al. 1989; Castor, Dykema, \&
Klein 1991).  In the test problem, we calculate the structure of the
radiation field in an uniform medium irradiated by a central source
in cylindrical geometry.  The opacity $\chi=0.2$ per unit
distance, and is uniform.  The grid runs from 0 to 1 in $Z$, and 0 to 20 in
$R$, with the central source placed on the axis of symmetry at $R=0$ and
$Z=0.5$.  The optical depth from the central source to
the outer boundary thus varies from 0.1 to 4, so as to test the full
range of behavior in the optically thin to optically thick limits.
The two calculations differ in the treatment of
the central source.  In ALTAIR, it
is represented by a cylinder, one zone in radius, and two zones along the axis,
whereas in the current method it is represented by a point on the axis.  The
grid for both simulations is 80 zones on a side.  The zone size varies, with
$\Delta Z$ and $\Delta R$ having minima of 0.01 at all of the boundaries
as well as at the central source, and increasing geometrically in between.
Variations in the grid size and spacing by factors of two had no effect on the
results.

Both here and in the results shown in \S~3, we take our boundary conditions for
the FLD calculation to
be reflection symmetry at $R=0$, and free-streaming at all other boundaries.
For the free-streaming condition, we use a one-dimensional Eddington
approximation.  To improve upon this would require knowledge of the angular
distribution of the radiation field, which is not available in the FLD method.
It is expected, therefore, to overestimate the flux through the boundary in
some regions in cases where the
scattered radiation is highly anisotropic, as when the source of scattering
is distant from, and has a grazing angle of incidence to the boundary point
in question.  This does not, however, appear to have a strong affect upon
our results.  We base this conclusion upon the good agreement, described
below, between results obtained with the current code and with ALTAIR
(which has information on the angular distribution of the radiation), the
good agreement between our disk models and earlier work (\S~3.2.1), and
the lack of sensitivity of the disk models to changes in the boundary
condition (\S~3.2.2).  This is perhaps not surprising.  In the cases where the
boundary flux calculation is least accurate, ie. where the radiation field is
dominated by scattered radiation, whose source has high inclination angle to
the boundary, the optical depth of the boundary cell in question is extremely
small.  In that case, the boundary condition finds a very weak vertical
gradient in $E\sr{sc}$, as would be expected, and the values of the energy
density are determined primarily by radial transport.  As a result, although
the actual calculations
of the fluxes at the boundaries in these regions are inaccurate, this
should not strongly affect the solution for the energy densities.

In Figure~1, we compare the total energy densities, $E_t=E\sr{dir}+E\sr{sc}$,
calculated by the current code and by ALTAIR for the test problem.  Results
from
ALTAIR are indicated by solid curves and those from the method described above
are indicated by dotted curves.  In Figure~1a are shown contours of $E_t$,
while slices of $E_t$ as a function of $R$ at $Z=0.5$ and $Z=0.01$ are
shown in Figures~1b and 1c, respectively.  Due to the linearity of the problem,
we are free to arbitrarily normalize the results of the two codes, which we
have done such that they agree at approximately
$R=1$, and $Z=0.5$.  As can be seen from the figure, the correspondence is
excellent, with the results of the two codes agreeing to within 50\%
throughout the computational grid.

The agreement is especially impressive
given the difference in computation times.  Due to the difficult nature of
the problem, the ALTAIR computation required $\approx10^4$ rays in order to
achieve convergence over a wide range of angles.  Even so,
slight inaccuracies due to a lack of angular resolution are apparent near
$\theta=0$ and $\theta=\pi/2$.  Elsewhere, however, the calculation has
converged, and further increases in the number of rays does not alter the
results.  The
large number of rays led to the calculation taking approximately 5 hours
on a Cray YMP.  By contrast, the method described above required
approximately 10 seconds, during which three iterations on the energy
density were performed.
Some of the difference in speed between the two codes can
be attributed to their different approaches to optimization.  ALTAIR is
optimized
such that it is vectorized over multiple frequency calculations.  The fact that
only one frequency is used in the current gray scattering problem then means
that ALTAIR is not being used in the most efficient manner.  Even allowing for
this, however, the difference in timing is substantial, and the speed and
accuracy of the
current method makes it feasible to include radiative transfer
effects in time-dependent hydrodynamics, which we will report in
subsequent work.

\yyskip
\centerline{\bf 3. ISOTHERMAL CORONAL MODELS}
\yyskip

\centerline{3.1. The Models}
\yyskip

As a first application of the method described in \S~2, we have modelled
static, isothermal accretion disk coronae in systems with luminosities
up to $L=0.59L\si{Edd}$, where $L\si{Edd}=4\pi G\mu_ec/\sigma_T$ is the
Eddington luminosity, and $\mu_e$ is the mean mass per electron in the gas.
Similar models
by OMK were limited to lower luminosities by the assumption that the
radiation field consisted only of direct and singly scattered radiation.
The accuracy of OMK's results in that limit, however, allow them to be used
as an additional test
of the current method at low luminosities, while the higher luminosity
models presented here form an extension of their work.
We summarize here the key features of the models of OMK, and refer the
reader to that work for further details.  The geometry of the models is
indicated schematically in Figure~2.
!!!!

The temperature of gas exposed to
a hard spectrum is determined by the ionization parameter, defined here
as the ratio of radiation-to-thermal pressure, $\Xi\equiv 4\pi J/cp$.
The resulting equilibrium temperature distribution is shown in Figure~3 for
two values of the characteristic energy of the radiation.  For
sufficiently hard spectra, a region of ionization parameters,
$\Xi_h<\Xi<\Xi_c$, exists for which the temperature is multiple-valued
(Krolik, McKee \& Tarter 1981).  For $\Xi>\Xi_h$, a high temperature
equilibrium exists in which Compton heating is balanced by Compton cooling,
and to a lesser extent bremsstrahlung emission.  For $\Xi<\Xi_c$ a low
temperature phase exists in which photoionization heating balances cooling
by line emission.  For $\Xi_h<\Xi<\Xi_c$ an intermediate, unstable
equilibrium temperature also exists, in which bremsstrahlung cooling
dominates.  By balancing Compton heating with bremsstrahlung and Compton
cooling, it is found that
$$\Xi_h=1.22T_{m8}^{-3/2}.\eqno{(3.1)}$$
In the above, $T_m=10^8T_{m8}$ is the Compton temperature of the total
radiation field, ie. the equilibrium temperature of gas under the action of
only Compton heating and cooling, defined by
$$T_m={{\sum\nolimits_iJ_iT_i}\over{\sum\nolimits_iJ_i}},\eqno{(3.2)}$$
where the sum runs over all radiation sources (direct and scattered
radiation from the central source plus disk radiation),
$T_i=\langle\varepsilon_i\rangle/4k$, and $\langle\varepsilon_i\rangle$ is
the energy-weighted mean photon energy of each source.

The multiple-valued region of the $\Xi-T$ equilibrium leads to a two-phase
structure vertically in accretion disks which are exposed to hard
radiation from a central source.
The decrease in gas density with increasing $Z$ leads to an increase in
$\Xi$.  For some $\Xi>\Xi_h$, the gas enters the hot phase, forming a
corona.  Allowing for thermal conduction from the hot gas to the cold gas,
McKee \& Begelman (1990) found that this occurs for
$$\Xi_b=1.06\Xi_h,\eqno{(3.3)}$$
which then defines the base of the corona.  Both here and below, quantities
defined at the base of the corona are indicated by
the subscript $b$.  The equilibrium temperature at $\Xi_b$ is determined by a
balance between Compton heating, and Compton and bremsstrahlung cooling.
This gives $T_b=0.47T_{m,b}\approx0.5T_{m,b}$, ie. about half the maximum
temperature attained in the absence of bremsstrahlung cooling.

The coronae are assumed to be
static.  In practice, this limits the models to radii less than a few tenths of
the Compton radius, defined by
$$\ric\equiv{{GM\mu}\over{k\tc}}={{9.81\times10^9}\over{\tic8}}
\left({M\over{M_\odot}}\right){\ \rm cm},\eqno{(3.4)}$$
where $M$ is the mass of the central object, and
$\tc=10^8 \tic8{\ \rm K}$ is the Compton temperature of radiation
from the central source.
As defined, $\ric$ is the radius beyond which the
sound speed of gas at temperature $\tc$ exceeds the escape velocity.  In
fact, substantial winds and mass loss are found at radii down to
about 0.1$\ric$ (Begelman, McKee \& Shields 1983; Woods et al. 1994), and
we have limited our models to radii $\xi_d=R\sr{max}/\ric=0.2$.  The Compton
radius forms a natural unit of distance, and our computations are
carried out using the dimensionless coordinates $\xi=R/\ric$ and
$\zeta=Z/\ric$.

In addition, the temperatures of the coronae
are assumed to be constant in the vertical direction (but not radially),
with a value
$$T_b(\xi)=T_{m,b}(\xi)/2,\eqno{(3.5)}$$
approximating the value found for the base of the corona by McKee \& Begelman
(1990).  The value of $T_m$ includes contributions from
direct, scattered, and disk radiation.  Using equation~(2.1) in (3.3) gives
$${{T_m}\over{\tc}}={{f\sr{dir}+f\sr{sc}+f_d(T_d/\tc)}\over f},
\eqno{(3.6)}$$ (OMK)
where we have assumed that the Compton temperature of the scattered radiation
$T\sr{sc}=\tc$.  The actual temperature throughout
the first scale height of the corona, $H_c(R)=\ric h_c(\xi)$,
which dominates the scattering, is within a factor
of two of that given by equation~(3.5), validating the usefulness of the
isothermal approximation.

The disk contribution is given by (Begelman \& McKee 1983)
$$f_d\equiv{{4\pi J_d}\over{L/4\pi R^2}}\equiv{{\xi_{hc}}\over\xi}=
{{4.5\times10^{-4}\tic8}\over{\epsilon_{-1}\xi}},\eqno{(3.7)}$$
and
$$T_d(\xi)=2.6\times10^4\left({L\over{L\si{Edd}\epsilon_{-1}}}\right)^{1/4}
\left({M\over{M_\odot}}\right)^{-1/4}\left({{T_{IC8}}\over\xi}\right)^{3/4}
{\ \rm K},\eqno{(3.8)}$$
where the luminosity efficiency
$\epsilon=0.1\epsilon_{-1}\equiv L/(\dot Mc^2),$
and where the mean intensity of the disk radiation exceeds that of the central
source for $\xi<\xi_{hc}$.  So long as $T_d(\xi_{hc})\ll T_{IC}$, then,
$\xi_{hc}$ represents the inner radius for which a substantial hot corona can
exist.  In terms of the Schwarzschild radius, $R\sr{S}=3\times10^{-5}\tic8
R\si{IC}$, we have $R_{hc}/R\sr{S}=15/\epsilon_{-1}$; for example, in an
X-ray binary containing a neutron star, $R_{hc}$ is several times the radius
of the neutron star.  From above, we have
$$T_d(\xi_{hc})=8.4\times10^6\left({L\over{L\si{Edd}}}\right)^{1/4}
\left({M\over{M_\odot}}\right)^{-1/4}\epsilon^{1/2}_{-1}.\eqno{(3.9)}$$
In most of the models shown below, we take
$M\approx1{\ \rm M}_\odot$, appropriate to X-ray binaries.  We then
have $T_d(\xi_{hc})/T_{IC}\approx0.07\tic8^{-1}$ for
$L/L\si{Edd}\approx0.5$.  The cooling effect of the disk is thus somewhat
limited, especially if the luminosity is high and $T_{IC8}<1$.  In low
mass systems, such as X-ray binaries, then, the disk may contribute
strongly to determining the coronal structure at small radii, as will be shown
below.  For larger values of $M$, the effect of disk cooling in quenching
the formation of a corona is much more pronounced.  For
example, if $M=10^8$~M$_\odot$, as may be appropriate for active galactic
nuclei, $T_d(\xi_{hc})\lapprox10^5$~K for all luminosities.  We examine the
effects of this in one model.

In the isothermal approximation, the density of the gas is given by
$$n_e(\xi,\zeta)=n_{e,b}(\xi){\rm exp}\left({-{{\tc}\over{T_b(\xi)}}}
\left\{\left[
\xi^2+h_d^2(\xi)\right]^{-1/2}-\left[\xi^2+\zeta^2\right]^{-1/2}\right\}
\right),\eqno{(3.10)}$$ (OMK),
where $h_d(\xi)=H_d(\xi)/\ric$ is the dimensionless
height of the disk.  The dimensionless scale height of the corona is
$$h_c(\xi)=\left[{{2T_b(\xi)}\over{\tc}}\right]^{1/2}
\xi^{3/2}.\eqno{(3.11)}$$
The electron density at the base, $n_{e,b}$, is
determined by both the radiation field and $\Xi_b$,
$$n_{e,b}={{4\pi J_b}\over{c\Xi_bkT_b}}{{n_e}\over{n\sr{tot}}}.\eqno{(3.12)}$$
The electron density is directly proportional to the
opacity, which we write in dimensionless form as
$$\tilde\chi(\xi,\zeta)\equiv\chi(\xi,\zeta)R\si{IC}=n_e(\xi,\zeta)\sigma_T
\ric.\eqno{(3.13)}$$

The models are computed on a nonuniform grid, in which $\Delta\xi$ and
$\Delta\zeta$ increase geometrically.  The minimum values are determined
such that $\xi_{hc}$, and $h_c(\xi_{hc})$ are resolved.  Because the
structure of the corona and the underlying disk are affected by inner
boundary conditions and by relativistic effects at radii less than
$\xi_{hc}$, we have simply assumed that the corona is negligible for
$\xi<\xi_{hc}$.  Including regions
with $\xi<\xi_{hc}$ would also be inconsistent in simulations of systems such
as black holes, for which the inner disk actually represents the central
source, but which our models treat as a point.  Because $\xi_{hc}$
forms the inner limit at which a substantial hot corona can exist, our choice
of resolution for $\xi<\xi_{hc}$ should have little effect upon the results.
Indeed, reductions in $\Delta\xi\sr{min}$ and
$\Delta\zeta\sr{min}$ by an order of magnitude was found to change the results
for $\xi>\xi_{hc}$ by only about 10\%.

Each of our sets of models includes a range of luminosities.
We begin at a low luminosity, for which the analytical results of
OMK can be used as a first estimate for $\tilde\chi(\xi,\zeta)$.  The opacity
distribution is used to determine $f\sr{dir}(\xi,\zeta)$, and
$f\sr{sc}(\xi,\zeta)$.  A new opacity distribution is then calculated using
equations~(3.1), (3.3) and (3.5)-(3.13), and the process is repeated until the
opacity distribution converges such that
$\left({{\delta\chi}\over\chi}\right)\sr{max}\le10^{-3}$.  The converged
distribution is then used as an
initial guess for a higher luminosity, and the entire process
is repeated for the new luminosity.

The models we have computed are listed in Table~1.  The parameters whose
effects we have examined are $\tc$, the disk flare, the source height, the
disk albedo, and the mass of the central object.  Model~1 is chosen to provide
a comparison with the numerical results of OMK, and we take it as our canonical
model.  In Model~2 we examine the effects of a realistic albedo for the disk,
rather than treating it as a black absorber.  Model~3 treats the case of a
source in the plane, with a flaring disk, while Models~4 and 5 examine the
two source/disk geometries at lower values of $\tc$.  Finally, Model~6 examines
the case of a massive central object.

\yyskip
\centerline{3.2. Results}
\yyskip

\centerline{\it 3.2.1. Comparison with Earlier Work}
\yyskip

Shown in Figure~4 are the results of Model~1, for which $\tic8=1$,
$h_d(\xi)=0$,
and $M=1$~M$_\odot$.  The source is placed along the symmetry axis, above
the plane a dimensionless distance $h_x=H_x/\ric=10^{-4}$.
As in the test case, $\xi=0$ is the axis of symmetry, and radiation is assumed
to free stream from the other boundaries, including $\zeta=0$ which
represents the surface of the disk.  By assuming that the X-rays free stream
at the lower boundary, we are therefore assuming zero albedo for the disk,
ie. it is assumed to be
a black absorber.  These model parameters were used by OMK as approximations
to the case of an extremely thin disk, with a central source raised above the
plane such that $h_x\gapprox[h_d(\xi_d)/\xi_d]^3$.  The same criterion applies
at much higher optical depths than could be considered by OMK, so long as
$f\sr{dir}+f\sr{sc}\approx const.$, as is found to be the case below.
Such raised sources are expected
from accretion onto highly magnetized neutron stars.  It may also occur for
accretion onto black holes, if the central disk thickens, or if much of
the emission is scattered from an optically thick jet.  Each of these
mechanisms results from emission from regions several
Schwarzschild radii in size (Takahara, Rosner, \& Kusunose 1989;
Begelman 1985).

In Figures~4a and 4b, we show $f\sr{dir}(\xi,0)$ and $f\sr{sc}(\xi,0)$,
respectively.  Figure~4c shows the coronal temperature, resulting from the
combination of direct, scattered, and disk radiation.  Apparent in the
figure is the initial decrease of $T_{m,b}$ with decreasing $\xi$ due to
the increasing dominance of disk radiation.  The eventual increase in
$T_{m,b}$ toward smaller $\xi$ is due to the increase in $T_d$.
Figure~4d shows the radial dependence of the characteristic optical depth,
$$\tau\sr{ch}(\xi)\equiv \tilde\chi(\xi,0)\xi,\eqno{(3.14)}$$
an indicator
of where the corona is optically thin in the radial direction.  In Figure~4e
is shown the perpendicular optical depth
$$\tau_\perp[\xi(j)]=\sum\limits_{k=1}^{k_{max}}\tilde\chi(j,k)\Delta\zeta(k).
\eqno{(3.15)}$$
The radiation in regions for which
$\tau_\perp\ll1$ is dominated by direct and singly-scattered radiation, for
which the analytical results of OMK are expected to be valid, and can be
used for comparison.  Finally, in Figure~4f are shown the dimensionless fluxes
incident upon the disk, which are important in determining the contribution
to line emission of fluorescence by the disk.  The dimensionless flux is
given by
$$f\si{F}(\xi,h_d)=f\sr{dir}(\xi,h_d)\cos(\theta)+f\sr{sc}(\xi,h_d)/\surd3,
\eqno{(3.16)}$$
where $\theta$ is
the angle between the ray from the source and the normal to the disk surface,
and the scattered contribution is calculated using the Eddington approximation.
Equation~(3.16) will overestimate $f_F$ in those regions where
$f\sr{sc}>f\sr{dir}\cos\theta$, and the corona is optically thin.  In that
case,
the scattered radiation may have a large inclination angle relative to the
disk.  Because the angular information of the scattered radiation is not
retained by the FLD approach, the estimate of equation~(3.16) cannot be
improved upon here, and we note that the values shown for $f_F$ should be taken
to be upper limits in those regions where $f\sr{dir}\cos\theta<f\sr{sc}$, and
$\tau\sr{ch}<1$.  The quantities in Figure~4 are each shown for values of
$L/L\si{Edd}$ of 0.064, 0.112, 0.194, 0.339, and 0.590.  Larger values could be
computed, but would be of questionable validity due to the dynamical
effects of
radiation pressure, not included in the current static calculations.  Contours
of the energy density, combined with vectors indicating the direction of the
radiation flux (not including $J_d$) are shown in Figure~5 for
$L/L\si{Edd}=0.590$.  The axes are the indices of the grid cells.  Because
$\Delta\xi$ and $\Delta\zeta$ increase geometrically outward, this emphasizes
the structure at small $\xi$ and $\zeta$.

The results for the two lowest luminosities shown in Figure~4 overlap the
luminosity range modelled by OMK, and so can be used as comparisons.  The
most direct comparison is given by the models with $L/L\si{Edd}=0.064$, for
which the results of the earlier work are indicated by crosses in Figures~4a-e.
For radii
$\xi\lapprox3\times10^{-3}$ at this luminosity, it was found in the earlier
work that $\tau_\perp>0.3$, which was taken to be the limit at which the
single-scattering assumption could be applied.  For all larger radii,
our results for $f\sr{dir}$ and $f\sr{sc}$ differ from those of OMK by
$\Delta\log f\approx0.5$, while $log(f\sr{dir}+f\sr{sc})$ differs by
approximately 0.3.  The agreement for $\tau_\perp$ is similar, while
the results are closer for $T_{m,b}$ and $\tau\sr{ch}$.

To ensure that the differences between our results and those of OMK are due
to multiple scattering, and do not result from differences in the two methods,
we have
also compared our results with theirs for $L/L\sr{Edd}=0.016$.  At this
luminosity, single scattering dominates for all $\xi>\xi_{hc}$, and $f\sr{sc}$
is of the same order as $f\sr{dir}$.  The luminosity of the system is therefore
sufficiently low for $f\sr{sc}$ to be dominated by singly-scattered
radiation, while also being sufficiently high that $f\sr{sc}$ contributes
significantly to determining the coronal structure.  We find that our results
for both $f\sr{dir}$ and $f\sr{sc}$
agree with those of OMK to within 20\% at all radii, implying that the
differences at higher
luminosities are, indeed, due to the addition of multiply scattered radiation.

At high luminosities ($L/L\sr{Edd}\gapprox0.1$), therefore, the analytic
approximations
of OMK underestimate the optical depths by factors of typically at least two.
While a general analytic fit to our results is not so straightforward as in
the case of lower optical depths, a fit to
our results in Figure~4d can be made using the results of \S~3.1, or,
equivalently, equations~(29a) and (29b) of OMK.  This gives
$$\tau\sr{ch}=1.54{{f_b\tic8^{3/2}}\over\xi}
\left({{T_{m,b}}\over{\tc}}\right)^{1/2}
{L\over{L\sr{Edd}}}.\eqno{(3.17)}$$
For $\xi\gapprox10^{-3}$, we find $f_b\approx f\sr{dir}+f\sr{sc}\sim0.1$
to within about a factor of two.  Using this in equation~(3.17) gives
$$\tau\sr{ch}\approx0.15{{\tic8^{3/2}}\over\xi}
\left({{T_{m,b}}\over{\tc}}\right)^{1/2}
{L\over{L\sr{Edd}}}{\ \ }(L\gapprox0.1L\sr{Edd}),\eqno{(3.18)}$$
which agrees with the results of Figure~4d to within 50\%.  For
$\xi\lapprox5\times10^{-3}(\tic8/\epsilon_{-1}$, $f_b\approx f_d$, and we find
$$\tau\sr{ch}\approx7\times10^{-4}{{\tic8^{5/2}}\over{\epsilon_{-1}\xi^{2}}}
\left({{T_{m,b}}\over{\tc}}\right)^{1/2}
{L\over{L\sr{Edd}}}{\ \ }(L\gapprox0.1L\sr{Edd}),\eqno{(3.19)}$$
which again agrees with the results of Figure~4d to within about 50\%.  The
transition between the regions where $f_b$ is dominated by $f_d$ and by
$f\sr{dir}+f\sr{sc}$ occurs at $\xi>\xi_{hc}$.  This occurs because $\xi_{hc}$
is defined by comparing $f_d$ with the {\it unattenuated} direct flux,
whereas in fact the flux from the central source is strongly attenuated
before reaching the base of the corona.  A reasonable approximation to
$\tau\sr{ch}$ is given by simply taking the maximum value of the two results
above.  The coefficient in equation~(3.18) rises to approximately 0.5 for
low luminosities
($L/L\sr{Edd}\lapprox0.1$).  Qualitatively, then, the presence
of multiply scattered radiation can be thought of as incompletely compensating
for the attenuation of the direct and singly-scattered radiation.

At low luminosities, the behavior of $f\sr{sc}$ with $\xi$ is similar to that
found by OMK, a central rise, with a flattening toward larger $\xi$.  The
decrease in $f\sr{sc}$ at $\xi\gapprox10^{-2}$
occurs when the optical depth from the disk to the
outer edge is less than unity.  At the highest luminosities, the behavior
changes significantly.  At small $\xi$, $f\sr{sc}$ decreases with luminosity
approximately as $L^{-1/2}$.  It also shows a minimum at
$\xi\approx3\times10^{-4}$, and rises toward larger radii approximately
as $\xi^{1/5}$.

We can gain a qualitative understanding of the behavior of $f\sr{sc}$ at
high luminosities using simplifying assumptions about the coronal structure.
Because $\tau\sr{ch}\gg1$, the
direct radiation penetrates only a short distance into the corona.  A
significant fraction of the
radiation striking the surface of the disk is due to a combination of
photons which are scattered near $\zeta=h_c(\xi)$ and then transported
vertically downward as well as photons emitted by the disk itself.  For
simplicity, we neglect the contribution of radial diffusion; as we shall see
below, this assumption renders our treatment quite approximate.  With this
assumption, we treat the system as a slab, irradiated from below by
an isotropic radiation source, $f_d$, and from above by a source with a
given angle of incidence, $f\sr{dir}$ (Chandrasekhar 1950; Busbridge 1960).  In
the Eddington approximation, at high optical depths, it is then found that
$$J_b=\left[\left(1+\surd3\mu_0\right)\mu_0{F\over{4\pi}}+
\left(\tau_\perp+{1\over{\surd3}}\right)
J_d\right]\left(\tau_\perp+{2\over{\surd3}}\right)^{-1},\eqno{(3.20)}$$
where $F=L/4\pi R^2$, $\mu_0$ is the cosine of the angle of incidence of
the incoming radiation, $\tau_\perp$ is approximately the
vertical optical depth from $h_c$ to $h_d$, and $J_d$ is given by
equation~(3.7).

We approximate the opacity distribution of the corona as
$$\tilde\chi(\xi,\zeta)=\cases{\tilde\chi_b(\xi),&$\zeta\leq h_c$;\cr 0,
&$\zeta>h_c$,\cr}\eqno{(3.21)}$$
where $\tilde\chi_b$ is given by equations~(3.12) and (3.13).  The surface
of the corona is then given by equation~(3.11).  We then have
$$\eqalignno{\tau_\perp\approx\tilde\chi_b
h\si{c}&=\tilde\chi_b\left({{T_{m,b}}
\over{\tc}}\right)^{1/2}\xi^{3/2}=\tau\sr{ch}\left({{T_{m,b}}\over\tc}\right)
^{1/2}\xi^{1/2},&(3.22a)\cr
&=1.54\tic8^{3/2}\left({{T_{m,b}}\over\tc}\right)\left({L\over{L\sr{Edd}}}
\right){{f_b}\over{\xi^{1/2}}}\equiv{{Af_b}\over{\xi^{1/2}}},&(3.22b)\cr}$$
from equations~(3.11), and (3.14)-(3.16).{\footnote{$^\dag$}{\rm The same
result was obtained by OMK in their equation~(55), which has a typographical
error: $\Xi$ should be replaced by $\tilde\Xi=0.65\tic8^{-3/2}$.}}

The cosine of the angle of incidence of the incoming radiation is given by
$$\mu_0=\sin\phi,\eqno{(3.23)}$$
where $\phi$ is the angle between the ray from the central source and the
plane of the corona at $\xi$, or
$$\phi=\phi_c-\phi_s={{{\rm d}h_c}\over{{\rm d}\xi}}-\arctan\left(\xi^{1/2}-
{{h_x}\over\xi}\right)\approx{1\over2}\xi^{1/2}+{{h_x}\over\xi},\eqno{(3.24)}$$
where $\phi_c$ and $\phi_s$ are the angles of the coronal surface defined
by the scale height, and of
the ray with respect to the plane, and the last approximation applies for
$h_x\ll\xi\ll1$.

Rewriting equation~(3.20) in terms of the normalized fluxes, equation~(2.1),
and using the relations above, we find
$$f_b\approx\left[\left(1+\surd3\mu_0\right)\mu_0
+\left(A\xi^{-1/2}f_b+{1\over{\surd3}}\right)f_d\right]
\left(A\xi^{-1/2}f_b+{2\over{\surd3}}\right)^{-1},\eqno{(3.25)}$$
giving a quadratic equation for $f_b$.  For $\xi\lapprox\xi_{hc}$,
equation~(3.25) gives $f_b\approx f_d$, as expected.  At larger radii,
equation~(3.25) tends asymptotically towards the result
$f_b\sim{1\over2}(3\xi)^{1/2}/\left[\left(1+3A/2\right)^{1/2}+1\right]$. As
can be seen from Figure~4b, however, we actually find
$f\sr{sc}(\xi,0)\propto\xi^{1/5}$ at large $\xi$, a much shallower variation
than predicted by equation~(3.25).  The two results cross at $\xi\approx0.1$,
and the prediction of equation~(3.25) lies below the numerical results at
smaller $\xi$.  The lack of a good fit is probably due to the inherent
two-dimensional nature of the problem.  We have calculated the net flux into
a column due to radial diffusion, averaged over one scale height, and compared
it with the flux due to direct radiation at one scale height.  The two are
found to be of the same order of magnitude for most radii, indicating that
radial diffusion plays an important role in determining the coronal structure
This conflicts with the qualitative impression of Figure~5, in which the fluxes
seem to be dominated by the vertical component.  That impression is due,
however, to the nonlinear nature of the axes, which emphasizes the structure
at small $\zeta$, whereas most of the contribution toward the radial flux
comes from $\zeta\sim h_c(\xi)$.

The significance of radial diffusion may also partly explain the difference
between our results and those of London (1985), who, in contrast to our
findings, concluded that the outer disk was strongly shadowed by the inner
corona at high luminosities.  That work, however, assumed that the scattered
radiation was transferred only vertically, and so would underestimate the
outward transfer of scattered radiation.

\yyskip
\centerline{\it 3.2.2. Dependence on Disk Albedo}
\yyskip

The results above assumed an albedo for the disk of zero.  This approximation
is very good for photon energies $E\lapprox10$~keV, and $E\gapprox100$~keV.  At
lower energies, absorption by photoionization dominates, while at higher
energies, Compton energy losses dominate.  At intermediate energies, however,
the reflectivity of cold matter reaches as high as $\approx50$\% (Lightman
\& White 1988).  Our calculations are monochromatic, and so we are unable
to account for the change in spectrum of the reflected radiation.
It is, however, of interest to investigate the qualitative effects of a
nonzero albedo for the disk.  This done for Model~2,
which is identical to Model~1,
except that we have taken the disk albedo to be 0.3, approximately the value
found by Lightman \& White (1988) for energies near 10~keV.  The results
of the two sets of models are very similar.  The coronal
densities are increased by the nonzero albedo by at most 50\% for some
radii.  The density increases due to the nonlinear nature of the problem;
the nonzero albedo increases $f_b$, which in turn increases the coronal
density, which itself increases $f_b$.  Overall, however, reasonable values
of the albedo do not strongly affect the results.

\yyskip
\centerline{\it 3.2.3. Dependence on $h_d$}
\yyskip

In the absence of jets, or a thick inner disk, accretion onto black
holes or weakly magnetized neutron stars will be characterized by a
much smaller value of $h_x$ than used above.
If $h_x\approx h\sr{d,min}$, where $h\sr{d,min}$ is the minimum value of
$h_d$, then disk flaring plays a crucial role in determining the
coronal structure.  The models of Shakura \& Sunyaev (1973) can be used to
predict $h_d(\xi)$ for regimes dominated by different opacities and
pressures.  In regions where gas pressure dominates radiation pressure, we
have approximately
$$h_d(\xi)=h_{d0}\xi^{1+s},\eqno{(3.26)}$$
where $h_{d0}$ is in the
range $10^{-3}$ to 0.1, and $s$ = 1/8 or 1/20.  For the systems of interest
here, we have taken $h_{d0}=0.02$ and $s$ = 1/8.

At small radii, radiation pressure dominates over gas pressure, and it is
found that the disk scale height is essentially constant with radius, giving
$$h\sr{d,min}={1\over2}\xi_{hc}(L/L\si{Edd}).\eqno{(3.27)}$$
Obviously, we must have $h_x>h\sr{d,min}$ or else the central regions will cast
a shadow over most of the disk, preventing the formation of a corona, unless
some dynamical process can act (Murray \& Lin 1992).  To
approximate the case $h_x\approx h\sr{d,min}$,
we have taken $h_x=h\sr{d,min}=0$, and taken $h_d(\xi)$ given by
equation~(3.26) in calculating Model~3, whose
results are shown in Figure~6.  Results
for systems in which $h\sr{d,min}<h_x<[h_d(\xi_d)/\xi_d]^3$ should be
intermediate to those of Models~1 and 3.

At small radii, the thin disk prescription of Shakura \& Sunyaev (1973)
breaks down, due to the increasing deviation of the flow from the assumed
Keplerian form, and radial derivatives must be included in the equations
of momentum and heat flux (Paczynski \& Bisnovatyi-Kogan 1981; Abramowicz
et al. 1988).  For $L/L\sr{Edd}\approx0.1$ the resulting changes in the disk
structure are small.  For $L/L\sr{Edd}\approx1$ however, the disk scale
height is found to be larger than in the thin disk derivation by approximately
an order of magnitude, out to several tens of Schwarzschild radii
(Abramowicz et al. 1988).  At larger radii, the corrections become increasingly
less important, and the disk structure tends increasingly toward the thin
disk results.  Because the latter have smaller scale heights, the result
is that the inner disk no longer flares, ($s<0$), possibly leading to a
shadowing of the outer disk by the inner disk.  In that case, a corona
could only exist in the outer disk if radiation is scattered
onto it by a corona formed at small radii.  The change in the
disk structure with increasing luminosity is highly nonlinear.  Because of
this, and because all of our models have $L/L\sr{Edd}<1$, we will not
consider this possible complication in our models here, but note it as a
possible issue at luminosities $L\approx L\sr{Edd}$.

Despite the relatively large flaring in Model~3, rays from the central source
to the
outer disk penetrate the corona at radii $\xi<\xi_{hc}$.  Unlike the previous
models, then, our results here are somewhat sensitive to assumptions
regarding the behavior of the disk within $\xi_{hc}$.  Following the
discussion of \S~3.1, it is assumed that the corona does not
exist inside of $\xi_{hc}$.
The result is to reduce the optical depths to the outer disk relative to the
predictions of the analytic results of OMK by factor of from two to three
for $\xi\gapprox10^{-2}$, and
$L/L\si{Edd}\lapprox0.2$.  At smaller radii and higher luminosities, the
results of OMK increasingly underestimate $\tau$.

At small radii, rays to the disk travel along the disk surface, resulting
in high optical depths from source to disk.  As a consequence, both $f\sr{dir}$
and $f\sr{sc}$ decrease towards small $\xi$.  Because $f\sr{sc}$ decreases
more shallowly than does $f\sr{dir}$, we find $f\sr{sc}>f\sr{dir}$ for
$\xi\lapprox0.98~(L/L\sr{Edd})^2$.  The dropoff of both $f\sr{dir}$ and
$f\sr{sc}$ also means that $f_d\gapprox f\sr{dir}+f\sr{sc}$ out to
$\xi\sim0.01$,
much further than in Model~1.  The increased dominance of the disk radiation
reduces the temperature of Model~3 by about an order of magnitude relative to
that of Model~1 at small radii.  From Equations~(3.1), (3.3) and (3.12), we
see that $\tau_{ch}$ varies approximately as $T_b^{1/2}$, and indeed it is
reduced by a factor of three relative to Model~1 at small radii.  Because
$\tau_\perp\approx\tilde\chi_bh_c\propto T_b$, it is much more sensitive
to the temperature decrease, and we find an order of magnitude reduction of
$\tau_\perp$ for $\xi\lapprox0.01$ relative to Model~1.
At larger radii, the disk flaring reduces the source-to-disk optical depth to a
value lower than in Model~1.  The changes in $\tau_{ch}$ and $\tau_\perp$ are
therefore much smaller for $\xi\gapprox0.01$.

\yyskip
\centerline{\it 3.2.4. Dependence on $\tc$}
\yyskip

Figure~7 shows the results for Model~4, which is identical to Model~3,
except that $\tic8=0.1$.  The analytic results of OMK
suggest that both $\tau\sr{ch}$ and $\tau_\perp$ vary as $\tc^{3/2}$
(assuming constant $f_b$ in equation~3.18), while in regions where
$f_d\gg f\sr{dir}+f\sr{sc}$ they should vary as $\tc^{5/2}$ (from
equation~3.19, assuming constant $T_{m,b}/\tc$).

As in the case of Model~3, we find that the disk radiation is dominant
at small radii.  As a result, the coronal temperature is reduced to about
$10^6$~K for $\xi\lapprox0.08\left(L/L\sr{Edd}\right)^{1.25}$.  At such
low temperatures, the thermal structure of the corona no longer has a
two-phase structure as depicted in Figure~3, so our results for this
region are quite approximate.  In the inner disk, where the disk radiation is
dominant, we find that $\tau_{ch}$ and $\tau_\perp$ vary as $\tc^{5/2}$.
At larger radii, the temperature dependence is closer to linear.  This is
shallower than $\tc^{3/2}$ because of the increase in $f\sr{dir}+f\sr{sc}$
in Model~4 relative to Model~1.  As a result of the different temperature
dependences of the two regions, the break between them
is even more apparent in Figure~7 than in Figure~6.

In Figure~8 shows the results for Model~5, which
is identical to Model~1, except that again $\tic8=0.1$.  Similarly to the
difference between Models~1 and 3, rays to the disk at small radii in Model~5
do not travel along the disk surface, resulting in much smaller optical depths
relative to Model~4.  The result is to increase the coronal temperature,
which remains near $10^7$~K throughout the simulation.  The thermal structure
of the corona is, therefore, two-phase at small radii, such that the results
of Model~5 are a better representation of the structure than those of Model~4.

A similar variation of the optical depths is found between Models~1 and 5
as was found above for Models~3 and 4.  At
$\xi\gapprox10^{-3}$, we find $\tau$ varies approximately
as $\tc$.  Again, this is shallower than predicted by equation~(3.18) due
to the increase in $f_b$ (see equation~3.12).  At smaller radii, where the
disk radiation dominates, the dependence steepens to about $\tc^2$.  This
is shallower than $\tc^{5/2}$ due to the increase in $T_{m,b}/\tc$ at
small radii in Model~5 relative to Model~1.  The increase results from the
lower optical depths, which allows more direct and scattered radiation to
penetrate, enhancing the coronal density.

As a consequence of the lower optical depths, the results are better
represented by the analytic approximations of OMK than are those of Model~1.
We find that the predicted values of $\tau$ agree with those of Figure~8
to better than 20\% for $L/L\sr{Edd}\lapprox0.3,$ while
the difference rises to 50\% at higher luminosities.  The agreement is not
as close for $f\sr{sc}$, for which we agree with the analytic values to within
50\% for $\xi\gapprox0.01$, while the difference increases to factors of
two or three at smaller radii and $L/L\sr{Edd}\gapprox0.3$.

\yyskip
\centerline{\it 3.2.5. Dependence on $M$}
\yyskip

Figure~9 portrays the results for Model~6.  It is identical to Model~1,
except that the central mass is taken to be $M=10^8$~M$_\odot$, characteristic
of the values expected in AGN.  The increased central mass
leads to a decrease in $T_d(\xi)$ by two orders of magnitude.  From
equation~(3.17), we that this leads to a quenching of the corona
at radii where $f_d\gapprox f\sr{dir}+f\sr{sc}$.  Indeed, we see
that Models~1 and 6 differ appreciably only for $\xi\lapprox10^{-3}$, where
the disk radiation dominates.
At these small radii, the reduction in $h_c$ reduces the optical depth,
allowing much more direct radiation to penetrate.  The increased direct
radiation reaching the disk partially offsets the lower temperature of the
disk radiation, such that $T_{m,b}/\tc$ is reduced by less than an order of
magnitude in Model~6 relative to Model~1, rather than by the two orders of
magnitude expected in the presence of pure disk radiation.  The values of
$\tau\sr{ch}$ and $\tau_\perp$ are correspondingly reduced relative to Model~1,
as expected from the discussion of \S~3.1.

\yyskip
\centerline{\bf 4. OBSERVATIONAL IMPLICATIONS}
\yyskip

One direct observational effect of the corona is in the modification of the
optical spectrum of the underlying accretion disk, due to Compton scattering
of the optical photons.  The continuum spectrum will be altered
significantly whenever the energy exchange
of the photons is comparable to their initial energy, which occurs when
the Compton y-parameter exceeds unity,
$$y\equiv{{4kT}\over{m_ec^2}}{\er{max}}\left(\tau_\perp,\tau_\perp^2\right)
=3.37\times10^{-2}T_{m,b8}{\er{max}}\left(\tau_\perp,\tau_\perp^2\right)>1
\eqno{(4.1)}$$
(Rybicki \& Lightman 1979), where as usual we have taken $T_b=T_{m,b}/2$
(equation~(3.5)).  Our numerical results show that $f_b\sim0.1$
in the outer disk for $L/L\sr{Edd}\gapprox0.1$, whereas $f_b\sim f_d$ in the
inner disk, where $f_d\gapprox0.1$ (see equation~3.7).  Using equation~(3.18),
we then find that $\tau_\perp$ is typically given by the maximum of
$$\tau_\perp
\approx0.15{{\tic8^{3/2}}\over{\xi^{1/2}}}\left({{T_{m,b}}\over{\tc}}\right)
{L\over{L\sr{Edd}}},\eqno{(4.2)}$$ and
$$\tau_\perp\approx7\times10^{-4}{{\tic8^{5/2}}\over{\epsilon_{-1}\xi^{3/2}}}
\left({{T_{m,b}}\over{\tc}}\right){L\over{L\sr{Edd}}}.\eqno{(4.3)}$$
In the former case (which applies when $f_d\ll f\sr{dir}+f\sr{sc})$, we have
$\tau_\perp>1$ for $\xi\lapprox0.024\left(L/L\sr{Edd}\right)^2\tic8^3
\left(T_{m,b}/\tc\right)$, where we find
$$y\approx8\times10^{-4}{{\tic8^4}\over{\xi}}
\left({{T_{m,b}}\over{\tc}}\right)^3
\left({L\over{L\sr{Edd}}}\right)^2.\eqno{(4.4)}$$
Thus, for most systems, $y$ exceeds 1 only for $\xi<\xi_{hc}$.  At such
small radii, $f_d\gapprox f\sr{dir}+f\sr{sc}$, and the second expression
for $\tau_\perp$ above holds, giving (again assuming $\tau_\perp>1$)
$$y\approx1.7\times10^{-8}{{\tic8^{6}}\over{\epsilon_{-1}^2\xi^{-3}}}
\left({{T_{m,b}}\over{\tc}}\right)^3
\left({L\over{L\sr{Edd}}}\right)^2.\eqno{(4.5)}$$
If $\tic8\approx1$, then the highest luminosity systems may have $y\gapprox1$
for $\xi\lapprox10^{-3}$.  If, however, $\tic8\approx0.1$, then we find
$y<1$ at all radii.

Irradiation by either direct or scattered X-rays also results in line
emission from the disk, which may be compared with observations.
Following Begelman \& McKee (1983), we define $\eta_l$ as the efficiency
with which the radiation incident upon the disk is converted into an
emission line $l$:
$$\eta_l\equiv{{F_l}\over{F\sr{inc}}},\eqno{(4.6)}$$
where $F\sr{inc}=({\bf F}\sr{dir}+{\bf F}\sr{sc})\cdot$\^n is the flux
incident upon the disk.  The line luminosity produced in an annulus of
width d$r$ at radius $r$ is d$L_l=4\pi r{\rm d}rF_l$, counting both sides of
the disk.  In terms of the dimensionless flux, $f_F$, introduced in
equation~(3.16), the line luminosity is
$${{{\rm d}(L_l/L)}\over{{\rm d}{\rm ln}\xi}}=\eta_lf_F.\eqno{(4.7)}$$
In applying this equation, it must be borne in mind that both $\eta_l$ and
$f_F$ depend upon the incident spectrum, which will differ from the
observed spectrum if photoelectric absorption is important.  Furthermore,
there is some flexibility in choosing the range of frequencies over which
$F\sr{inc}$ is to be evaluated, and this has a corresponding effect on the
value of $\eta_l$.

As an example, consider the formation of the 6.4~keV Fe~K$\alpha$ fluorescence
line by the absorption of photons of energy $\epsilon>\epsilon_K=7$~keV
by cold gas in the disk.  Let $L_K$ be the luminosity per unit energy at the
K absorption edge.  Choose $L$ to be $\epsilon_KL_K$; the effects of the
shape of the spectrum on the line luminosity are then buried in the efficiency
factor $\eta_l$.  In terms of the equivalent width, $W_\epsilon$ given by
$L_l\equiv L_KW_\epsilon$, equation~(4.7) then becomes
$${{{\rm d}(W_\epsilon/\epsilon_K)}\over{{\rm d}{\rm ln}\xi}}=\eta_lf_F.
\eqno{(4.8)}$$
Detailed calculations are needed to determine the value of $\eta_l$, but
we expect it to be of order 0.1 since the fluorescence yield is about
1/3 for neutral iron (Krolik \& Kallman 1987), iron accounts for about half of
the total
cross section just above the K edge (e.g. Fireman 1974), and about half of
the fluorescence photons will escape from the disk without being absorbed.
Our calculated values of $f_F$ are upper limits to the true values in this
case, since they ignore photoelectric absorption, which can be comparable
to Compton scattering in coronal gas (Krolik \& Kallman 1987).

\yyskip
\centerline{\bf 5. CONCLUSIONS}
\yyskip

We have developed a computationally efficient method for calculating
two-dimensional monochromatic radiative transfer in accretion disk coronae.
The method splits the radiation into direct and scattered components.  The
direct radiation is calculated by integrating the optical depth along rays
from the central source, while transfer of the scattered radiation is
approximated by flux-limited diffusion.  The optically thin and thick limits
are therefore handled with great accuracy.  In fact, in tests against a
full 2D radiative transfer code, ALTAIR, the method is found to be reasonably
accurate in all optical depth regimes.
The efficiency of the code makes it practical for use in hydrodynamic
simulations of accretion disk coronae and winds.  Those simulations can
then be run for systems with luminosities $L\approx L\sr{Edd}$, allowing
a full exploration of the rich dynamic effects expected in systems at
high luminosities.  The results of this will be reported in future work.

As a first application of the radiative transfer method, we have computed
self-consistent static models of Compton-heated accretion disk coronae.
At low luminosity, these models have been compared with the results of earlier
work by Ostriker, McKee, \& Klein (1991), which included only direct and
singly-scattered radiation.  The
models are found to agree well with the earlier work in all regimes where
multiply-scattered radiation is expected to be unimportant.

At high luminosities ($L/L\sr{Edd}\gapprox0.1$), where multiply-scattered
radiation dominates the
radiation field, we find that our results form a roughly continuous
extension of the results at lower luminosities.  Qualitatively, then, it
appears
that, rather than dramatically increasing or decreasing the irradiation of
the outer disk, the multiply-scattered radiation acts to compensate partially
for the loss of
direct and singly-scattered radiation as the luminosity increases.  In fact,
we find that the normalized scattered intensity, $f\sr{sc}$, varies relatively
little over a wide range of source luminosities.  This
continuity with the low luminosity results is contrary to what might have
been expected with the inclusion of multiple scattering.  It also disagrees
with some earlier results (London 1985), which indicated that shadowing from
the inner corona can prevent the formation of extensive coronae at high
luminosities.  Some of the difference may be due to neglect of the cooling
effect of the disk radiation in the earlier work, which reduces the optical
depth of dense gas at small $R$.  More importantly, it was also assumed by
London (1985), that the transfer of scattered radiation could be approximated
as one-dimensional in the vertical direction.  We find, however, that
analytic results derived using this assumption do not form good approximations
to our results.  The reason is that radial diffusion, averaged over the height
of the corona, is at least as important as vertical transport.  The problem
is, therefore, inherently two-dimensional in nature.

We have estimated the observational consequences of our coronal models both
for coronal veiling, and fluorescent line emission.  In general, it is found
that the Compton $y$ parameter is less than unity.  Values $y\sim0.1$
are, however, attained, which may have measurable effects upon line
structures.

To estimate the fluorescent line emission, we take as
an example Model~1, for which $f_F\sim0.05$ for $\xi\lapprox0.1$, and varies
only slightly with luminosity.  Using
equation~(4.8) with $\eta_l\sim0.1$, we find $W_\epsilon\sim100$~eV for the
6.4~keV fluorescence line.  This is similar to the values found by
White et al. (1986) for the low mass X-ray binary systems GX17+2, X1705-440,
Ser X-1, and XB1728-337.  The
emission spectra to be expected from irradiated accretion disks in our models
will be examined in future work.

\yskip
The authors would like to acknowledge helpful discussions with T. Kallman,
R. London, R. Rogers, G. Rybicki, J. Wilson, and D. T. Woods.  We would also
like to thank
P. Dykema for providing the ALTAIR code used as a test.  The research of RIK,
JIC, CFM, and SDM is supported in part by NASA grant NAGW-3027 to the High
Energy Astrophysical Theory and Data Analysis Program.  The computations were
carried out on the YMP 8/128 supercomputer at LLNL.  This work was performed
in part under the auspices of the U.S. Department of Energy at the Lawrence
Livermore National Laboratory under contract W-7405-ENG-48.  The authors
also gratefully acknowledge the publication policy of I. Stravinski.

\par\vfill\eject
\centerline{\bf References}
\yyskip

\pp Abramowicz, M. A., Czerny, B., Lasota, J. P., \& Szuszkiewicz, E. 1988,
ApJ, 646

\pp Begelman, M. C. 1985, in {\it Astrophysics of Active Galaxies and
Quasi-Stellar Objects}, J. S. Miller ed. (Mill Valley: University Science
Books), p. 411

\pp Begelman, M. C., \& McKee, C. F. 1983, ApJ, 271, 89

\pp Begelman, M. C., McKee, C. F., \& Shields, G. A. 1983, ApJ, 271, 70

\pp Bowers, R. L., \& Wilson, J. R. 1991, {\it Numerical Modeling in
Applied Physics and Astrophysics}, (Boston: Jones and Bartlett), p. 418

\pp Busbridge, I. W. 1960, {\it The Mathematics of Radiative Transfer},
(Cambridge: Cambridge Univ. Press)

\pp Castor, J. I., Dykema, P. G., \& Klein, R. I. 1991, in {\it Stellar
Atmospheres: Beyond Classical Models}, ed. L. Crivellari et al.,
(Dordrecht: Kluwer), p. 49

\pp Chandrasekhar, S. 1950, {\it Radiative Transfer}, (Oxford: Clarendon Press)

\pp Fireman, E. L., 1974, ApJ, 187, 571

\pp Kallman, T. R. 1990, in {\it Accretion-Powered Compact Binaries}, ed.
C. W. Mauche (Cambridge: Cambridge Univ. Press), p. 325

\pp Kershaw, D. S. 1978, {\it J. Comp. Phys.}, 26, 43

\pp Klein, R. I., Castor, J. I., Greenbaum, A., Taylor, D., \& Dykema, P. G.
1989, JQSRT, 41, 199

\pp Ko, Y-K., \& Kallman, T. R. 1993, in {\it The Evolution of X-ray
Binaries}, in press

\pp Krolik, J. H., \& Kallman, T. R. 1987, ApJL, 320, L5

\pp Krolik, J. H., McKee, C. F., \& Tarter, C. B. 1981, ApJ, 249, 422

\pp Levermore, C. D. \& Pomraning, G. C. 1981, ApJ, 248, 321

\pp Lightman, A. P. \& White, T. R. 1988, ApJ, 335, 57

\pp London, R. A. 1985, in {\it Cataclysmic Variables and Low-Mass X-Ray
Binaries}, D. Q. Lamb and J. Patterson eds. (Dordrecht: D. Reidel), p. 121

\pp McKee, C. F. \& Begelman, M. C. 1990, ApJ, 358, 392

\pp Murray, S. D. \& Lin, D. N. C. 1992, ApJ, 384, 177

\pp Ostriker, E. C., McKee, C. F., \& Klein, R. I. 1991, ApJ, 377, 593 (OMK)

\pp Paczynski, B., \& Bisnovatyi-Kogan, G. 1981, {\it Acta. Astronomica},
283

\pp Rybicki, G. B., \& Lightman, A. P. 1979, {\it Radiative Processes in
Astrophysics} (New York: Wiley-Interscience), p. 208

\pp Shakura, N. I., \& Sunyaev, R. A. 1973, A\&A, 24, 337

\pp Takahara, F., Rosner, R., \& Kusunose, M. 1989, ApJ, 346, 122

\pp Tuchman, Y., Mineshige, S., \& Wheeler, J. C. 1990, ApJ, 359, 164

\pp White, N. E., Peacock, A., Hasinger, G., Mason, K. O., Manzo, G., Taylor,
B. G., \& Branduardi-Raymont, G. 1986, MNRAS, 218, 129

\pp Woods, D. T., Klein, R. I., Castor, J. I., \& McKee, C. F. 1994,
ApJ, submitted

\par\vfill\eject

\vbox{%
\centerline{{\twelvecaps TABLE 1\ }}
\centerline{The Models}
\vskip 10pt
\centerline{\vbox{\halign {\hfil #\hfil &
\quad \hfil #\hfil & \quad \hfil # \hfil & \quad \hfil #\hfil & \quad \hfil #
\hfil & \quad \hfil #\hfil \cr
\dline
Model & $T\si{IC8}$ & flare$^a$ & $h_x$ & A$^b$ & M$\sr{cent}$/M$_\odot$ \cr
\cline
1  &  1.0 & n & 10$^{-4}$ & 0.0 & 1.0    \cr
2  &  1.0 & n & 10$^{-4}$ & 0.3 & 1.0    \cr
3  &  1.0 & y & 0.0       & 0.0 & 1.0    \cr
4  &  0.1 & y & 0.0       & 0.0 & 1.0    \cr
5  &  0.1 & n & 10$^{-4}$ & 0.0 & 1.0    \cr
6  &  1.0 & n & 10$^{-4}$ & 0.0 & 10$^8$ \cr
\bline
}}}}

\pp{$^a$}n = flat disk, y = flaring disk (equation~3.26 with $h_{d,0}=0.02$
and$s=1/8$).

\pp{$^b$}Disk albedo.

\vfill\eject

\centerline{\bf Figure Captions}
\yyskip

\noindent{\bf Figure 1.} Test case comparing the results of the combined ray
tracking/flux-limited diffusion code described in the text (dotted curves) with
those of the two-dimensional version of ALTAIR (solid curves).  The system
has uniform opacity, $\chi=0.2$ per unit distance, distributed in a disk
with dimensions 0 to 1 in $Z$, and 0 to 20 in $R$.  The energy densities from
the two codes have been arbitrarily normalized such that they cross near
$(R,Z) = (1,0.5)$.  Figure~1a shows contours of the
total energy density.  The contour spacing is 0.5 dex.  Figure~1b
shows the energy density as a function of radius at $Z=0.5$, while Figure~1c
shows the same quantities for $Z=5\times10^{-3}$.

{\noindent\bf Figure 2.} Schematic plot indicating the geometry of the models.
Only the innermost regions are shown for clarity.  The actual models extend
to $\xi=\zeta=0.2$.
Shown are the disk surface, $h_d(\xi)$, (long-dashed curve), the scale height
of the corona, $h_c(\xi)$, (short-dashed curve), and a sample ray
from the central source to the disk surface (solid curve).
Also indicated by the horizontal and vertical
dotted curves are the values of $h_x$ and $\xi_{hc}$, respectively.  We use
$h_x=3\times10^{-5}$, rather than 10$^{-4}$ as used in the models, again for
clarity of presentation.  The value
of $\xi_{hc}$ assumes $\tic8=1$.  In the models described in \S~3, either
$h_x=0$,
or $h_d(\xi)=0$.  In the former case, $h_d(\xi)$ forms the lower boundary
of the computational grid, and rays from the central source are concave
downward on the computational grid.

{\noindent\bf Figure 3.} Equilibrium temperature as a function of ionization
parameter for radiation fields characterized by Compton temperatures of
10$^8$~K (solid curve), and $1.5\times10^7$~K (dashed curve).  The
quantities $T_m$, $T_b$, $\Xi_h$, $\Xi_b$, and $\Xi_b$ are indicated for
the 10$^8$~K curve, while the curve for $1.5\times10^7$~K illustrates the
variation of these quantities with $T_m$.

\noindent{\bf Figure 4.} Fluxes and opacities calculated at the base of the
corona are
shown as a function of dimensionless radius $\xi=R/R\si{IC}$ for models with
$\tic8=1$, $h_d(\xi)=0$, and zero disk albedo.  Results are shown for five
different luminosities,
$L/L\si{Edd}$ = 0.064, 0.112, 0.194, 0.339, and 0.590, as indicated by the
solid, short dashed, medium dashed, long dashed, and long/short dashed
curves, respectively.  Figures~4a and 4b show the normalized fluxes
$f\sr{dir}\equiv e^{-\tau}$, and $f\sr{sc}$, respectively, while Figure~4c
shows
the ratio $T_{m,b}/\tc$.  The actual coronal
temperature is taken to be $T_{m,b}/2$.  The characteristic optical depth,
$\tau\sr{ch}=\tilde\chi\sigma\si{T}\xi$, is shown in Figure~4d, while the
optical depth perpendicular to the disk is shown in Figure~4e.  Finally,
in Figure~4f are shown the dimensionless fluxes incident upon the disk.  The
crosses in Figures~4a-e indicate the results of Ostriker et al. (1991) for
$L/L\si{Edd}=0.064$.  As can be seen, our results for that luminosity differ
from the earlier work by approximately 0.5 in the logarithm for $f\sr{dir}$,
$f\sr{sc}$, and $\tau_\perp$, with better agreement for $T_{m,b}/\tc$ and
$\tau\sr{ch}$.  That the difference is probably the result of the inclusion of
multiply-scattered radiation in the current method is indicated by the
agreement of the two methods to better than 20\% at lower luminosities.

{\noindent\bf Figure 5.} Contours of the total energy density from the model
with $L/L\si{Edd}=0.59$ of Figure~4 are shown.  Superimposed upon these are
vectors
indicating the direction of the radiative flux.  The magnitude of the flux
is not indicated, due to its large dynamic range.  Rather than the real
coordinates, the axes are the indices of $R$ and $Z$.  Because of the geometric
increase in the zone spacing, this effectively expands the inner regions of
the solution.

{\noindent\bf Figure 6.} The same as Figure~4, for Model~3, in which
the disk flares as described in the text.  The slight kink in $f\sr{dir}$
for $L/L\sr{Edd}=0.064$ is not physical, but varies with the spatial
resolution.

{\noindent\bf Figure 7.} The same as Figure~6, for Model~4, in which
$\tic8=0.1$.

{\noindent\bf Figure 8.} The same as Figure~4, for Model~5, in which
$\tic8=0.1$.

{\noindent\bf Figure 9.} The same as Figure~4, for Model~6, in which
$M=10^8$~M$_\odot$.

\par\vfill\eject
\end